\documentclass[preprint,showpacs,preprintnumbers,amsmath,amssymb,superscriptaddress]{revtex4}

\usepackage{graphicx}

\usepackage{bm}

\newcommand{\be}[1]{
\begin{equation} \label{#1}}
\newcommand{\ee}{\end{equation}}
\newcommand{\ba}{\begin{eqnarray}}
\newcommand{\ea}{\end{eqnarray}}

\newcommand{\pderiv}[2]{\frac{\partial #1}{\partial #2}}

\newcommand{\deriv}[2]{\frac{d #1}{d #2}}

\begin{document}

\title{Surface-active dust in a plasma sheath}

\author{ A.M. Ignatov}\email{aign@fpl.gpi.ru}
\affiliation{General Physics Institute, Moscow, Russia}

\author{ P.P.J.M. Schram}\affiliation{Eindhoven University of Technology,
  Eindhoven,
  The Netherlands}
  \author{S.A. Trigger}\affiliation{Humboldt University,  Berlin, Germany}
\begin{abstract}
The  inhomogeneity of the plasma pressure near a conducting
electrode is a cause for introducing  the surface tension
associated with the plasma-electrode interface. We evaluate the
dependence of the surface tension on the density of the charged
dust immersed in the plasma sheath. In a wide range of parameters,
the surface tension turns out to be an increasing function of the
dust density.
\end{abstract}

\date{\today}

 \maketitle

 Throughout the recent  decade, an admixture of charged macro-particles, electrons, ions, and
neutral atoms, called a dusty plasma, has been the subject to some
thousands of studies.  Numerous  aspects of the  physics of dusty
plasmas are discussed in the series of reviews starting with
\cite{TMT02} and the monograph \cite{shukla02}.

 Many laboratory experiments deal with
relatively small number of dust grains levitating above a
horizontal  rf-powered or dc-biased  electrode. Under these
conditions, dust grains are negatively charged and the gravity
force  is compensated by the strong electric field in a plasma
sheath adjacent to an electrode. Dust suspended in a plasma sheath
self-organizes itself in various single or multi-layered
structures. In particular, recent experiments \cite{tue}
demonstrated that with growing rf power or gas pressure there
evolves a void, that is, a dust-free region, in the center of
 a single dust layer. Since negatively charged grains
 levitating at the same height repulse one another via the screened
 Coulomb potential, there are no evident reasons for the void emergence.
    The explanation of this phenomenon given in \cite{tue} implies
    existence of an attractive interaction between grains; however, the
    nature of the interaction remains obscure.

 To avoid confusion, it should be stressed that
    three-dimensional voids, which have been observed earlier
    \cite{goree,therm,mg1,PKE}, evolve in  the plasma bulk.
    The theory of three-dimensional voids \cite{theory1,theory2} demands
    large number of grains and strong influence of the dust
    upon the discharge structure. At first sight, both assumptions fail under the
    conditions of the experiment \cite{tue}, where some hundreds
    of grains only have been used.

 The main purpose of this Letter is to  introduce the concept of
 plasma surface tension. We demonstrate that  even
a small number of dust grains can modify the plasma sheath profile
in such a way that the sheath may confine the dust in the
horizontal direction. Our reasoning is based on the analysis of
the plasma pressure in the sheath-presheath area of a gas
discharge. We show that the horizontal electric  force appearing
due to the inhomogeneity of a dust layer is conveniently treated
in terms of the plasma surface tension.  It should be stressed
that explanation of
 the experiment  \cite{tue} is beside our purpose: there is
  evidently a number of relevant factors that are not discussed here.

\begin{figure}
\includegraphics{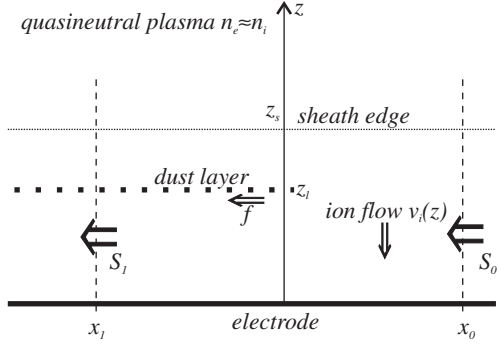}
 \caption{\label{fig1} Schematic
of the model }
\end{figure}
The schematic of the model implemented here is depicted in
Fig.~\ref{fig1}. The  plasma consisting of electrons and singly
charged ions is described with the help the continuity equation

\be{cont} \deriv{n_i(z) v_i(z)}{z} =\nu_{ion} n_e(z),\ee
the momentum balance equation

\be{mom} \deriv{m_i n_i(z) v_i^2(z)}z + e n_i(z) \deriv{\phi(z)}z
=0,\ee
and Poisson's equation

\be{poisson} \frac{d^2\phi(z)}{d z^2}= 4\pi e(n_e(z)-n_i(z))-4 \pi
\rho_d(z).\ee

Here, $\nu_{ion}$ is the electron impact ionization frequency and
$\rho_d(z)$ is the charge density of aerosol grains. Other
notation is standard. Electrons are supposed to be Boltzmannian:
$n_e(z)=n_0 \exp\left(e \phi(z)/T_e\right)$, where $n_0$ is a
normalizing constant. We also assume that the mean free path
 of the ion-neutral collisions is large compared to the Debye
 length, which determines the spatial scale of  the
sheath.

Since the electron and ion fluxes at the electrode ($z=0$) are
equal,  one of the boundary conditions to
Eqs.~(\ref{cont}-\ref{poisson}) is

\be{wall} n_i(0)v_i(0)+n_e(0) \sqrt{\frac{T_e}{2\pi m_e}}=0.\ee

Far from the electrode the plasma is quasi-neutral, \textit{i.e.,}
$n_e=n_i=n_0$. Generally, the equilibrium plasma density, $n_0$,
is the eigenvalue of the problem and it is determined by the
balance of ionization and losses of charged particles. For this
reason, the bulk plasma density may depend on the dust charge
located in the sheath because the latter modifies the electric
field profile, which, in its turn, influences the net particle
flux at the electrode. However, here we assume the  overall size
of the plasma to be large enough, so that the equilibrium density,
$n_0$, is conditioned by the recombination in the plasma volume.
Therefore, the plasma density tends to a certain fixed value at
$z\to \infty$. In order to patch the quasi-neutral plasma and the
sheath, the equations (\ref{cont},\ref{mom}) are solved assuming
that $n_e(z)=n_i(z)$ and $\rho_d(z)=0$ (\textit{e.g.,}
\cite{riemann}) that results in

\be{densN}n_i(z)=\frac{n_0 v_s^2}{v_s^2+v_i(z)^2},\ee where
$v_s=\sqrt{T_e/m_i}$ is the ion sound velocity.  It is assumed
that at the end of the simulation area, $z=z_0$, the ion flow
velocity is fixed, \textit{i.e.,} $v_i(z_0)=v_0$. The value of
$v_0$ should be small enough;  all solutions exemplified below are
obtained for $v_0=-0.01 v_s$. Then, Eqs. (\ref{cont},\ref{densN})
are used to evaluate boundary values of the density, $n_i(z_0)$,
the potential, $\phi(z_0)$, and its derivative,
$\phi^\prime(z_0)$; this results in four boundary conditions for
Eqs.~(\ref{cont}-\ref{poisson}). The length of the simulation
area, $z_0$, is determined with the help of the fifth boundary
condition (\ref{wall}).

In the framework of our model, the dust is represented as an
infinitesimally thin charged massive layer levitating at $z=z_l$,
that is, its charge density is

\be{rho} \rho_d(z)=-q \sigma \delta(z-z_l),\ee
where $-q$ ($q>0$) is the charge of a single grain and $\sigma$ is
the surface dust density. The equilibrium levitation height,
$z_l$, depends on the vertical force balance. The largest forces
acting upon dust in a plasma sheath are the electric field force
and the gravity force. Since the $z$-component of the electric
field, $E(z)=-\phi^\prime(z)$, is discontinuous at $z=z_l$, the
vertical force balance is written as

\be{vf} \frac12 q \left[E(z_l+0)+E(z_l-0)\right]=M g, \ee
where $M$ is the mass of a single grain and $g$ is the gravity
acceleration (in Fig.~\ref{fig1}, the Earth is at the bottom). On
the other hand, the electric field discontinuity is given by

\be{efd} E(z_l+0)-E(z_l-0)=-4 \pi q \sigma \ee
and the potential, $\phi(z)$, is a continuous  function at
$z=z_l$. The relations (\ref{vf},\ref{efd}) allows us to patch the
numeric solutions of Eqs.~(\ref{cont}-\ref{poisson}) below
($0<z<z_l$) and above ($z_l<z<z_0$) the layer, and to obtain the
equilibrium levitation height, $z_l$, and the net thickness of the
sheath-presheath area, $z_0$.

It is well-known that the grain charge, $q$, in
Eqs.~(\ref{vf},\ref{efd}) depends on the plasma parameters, and
its variability should be also taken into account \cite{vc00}.
However, here we assume that the grain charge is fixed; the
results of preliminary computer runs with the variable grain
charge did not alter our main conclusions. We also ignore the ion
wind force in Eq.~(\ref{vf}) as long as absorption of plasma by
dust grains. Simple estimations show that these processes are of
minor importance.

Our main objective is to analyze the plasma pressure profile in
the sheath-presheath area. Evidently,
Eqs.~(\ref{cont}-\ref{poisson}) ensure the conservation of the net
linear momentum of the system. The tensor of the momentum flux, or
the pressure tensor is a sum of the ion momentum flux, the
electron pressure and Maxwell's stress: $P_{\alpha\beta}=m_i n_i
v_{i\alpha}v_{i\beta}+n_e T_e \delta_{\alpha\beta}+
\boldsymbol{E}^2 \delta_{\alpha\beta}/8\pi-E_\alpha E_\beta /4\pi$
($\alpha,\beta=x,y,z$). In the one-dimensional model described
above, the pressure tensor is diagonal: $P_{zz}=P_l$ and
$P_{xx}=P_{yy}=P_{tr}$. The longitudinal pressure is

\be{pl} P_l(z)=m_i n_i(z) v_i(z)^2+ n_e(z)T_e-\frac{1}{8\pi}
E(z)^2.\ee

Since within our model there are no momentum losses in the plasma,
 $P_l$ is constant in the absence of dust and it is equal to the
plasma electron pressure at $z\to\infty$, $P_l=n_0 T_e$.

The transverse  pressure is written as

\be{pt} P_{tr}(z)=n_e(z)T_e+\frac{1}{8\pi} E(z)^2.\ee
In contrast with Eq.~(\ref{pl}), there is no contribution of the
ion momentum flux and the electric field pressure is of the
opposite sign.

Although the transverse pressure tends to the electron pressure in
the plasma bulk, $P_{tr}|_{z\to\infty}=n_0T_e$, it is essentially
nonuniform   in the sheath-presheath area adjacent to the
electrode. This circumstance allows us to define the
\textit{surface tension} at the plasma-electrode interface as

\be{ps} s(\sigma)=\int\limits_0^\infty dz\, \left(n_0 T_e -
P_{tr}(z)\right).\ee
The physical meaning of this quantity is fairly evident. Since the
bulk plasma pressure is positive, two parts of the discharge
repulse one another.  As we shall demonstrate later, the
transverse pressure (\ref{pt}) is reduced near the wall;
therefore, there arises   the additional horizontal force
attracting two parts of the plasma-wall interface. The surface
tension (\ref{ps}) characterizes the magnitude of this force per
unit length.  It should be noted that usually surface tension is
defined in terms of surface free energy. The latter is obviously
meaningless in application to an open, nonequilibrium system like
the plasma-wall boundary. However, due to the momentum
conservation, the surface tension given by (\ref{ps}) is a
well-defined concept. Of particular interest for our purposes is
the dependence of the surface tension (\ref{ps}) on the dust
density.

The utility of the notion (\ref{ps})  may be illustrated in the
following way. Suppose that the layer density is a weakly varying
function of the transverse coordinate, $x$, that is,
$\partial\sigma/\partial x \sim \epsilon $, where $\epsilon \ll
1$. Integrating $P_{xx}$ component of the pressure tensor with
respect to $z$ at nearby points $x$ and $x+\Delta x$ we see that
the horizontal force (\textit{i.e.,} the momentum flux) acting
upon a small part of the layer of the width $\Delta x$ is written
in two identical forms:

\be{f1} F_x=-q \sigma E_x \Delta x \equiv -\Delta x
\pderiv{}{x}\int dz\, P_{xx}(x,z).\ee

Comparing both parts of the identity and noting that
$E_x\sim\partial/\partial x\sim O(\epsilon)$ we conclude that in
order to obtain the first-order term of the expansion of the
horizontal force in powers of $\epsilon$ it is sufficient to
evaluate the surface tension (\ref{ps}) ignoring the spatial
inhomogeneity of the layer density.
 Therefore, $f_x=\partial s(\sigma(x))/\partial x$. This simple
 reasoning may be confirmed by the  more tedious asymptotic
 expansion of the hydrodynamic equations in powers of $\epsilon$.

We can also extract some useful information about the horizontal
force even for the sharply bounded layer.   Suppose that the dust
layer occupies the area $x<0$, as depicted in Fig.~\ref{fig1}. The
levitation height, $z_l$, now depends on $x$, the electric field
and the ion flow near the edge of the layer are essentially
two-dimensional. However,  the horizontal component of the force
is the difference $f_x=s_0-s_1$, where $s_{0,1}$ are given by the
integral (\ref{ps}) evaluated across the sheath in the dust-free
area, at $x=x_0$, and in the presence of dust, $x=x_1$
(Fig.~\ref{fig1}).
 Assuming that both points $x_{0,1}$ are far
enough from the layer edge, the horizontal components of the ion
velocity and the electric field are negligible. Therefore, the
values of the surface tension, $s_{0,1}$, which determine the net
force exerted upon the edge, are provided by the solution of the
one-dimensional problem (\ref{cont}-\ref{poisson}).

\begin{figure}
\includegraphics{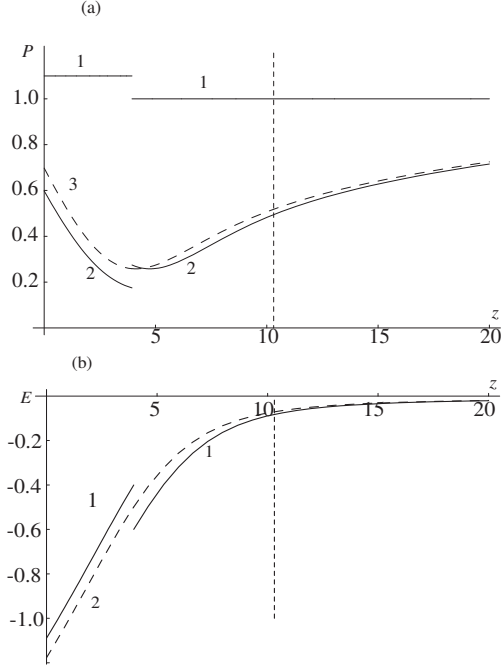}
 \caption{\label{fig2} The pressure (a) and the electric field (b) profiles in the sheath-presheath area evaluated
 for $\mu=0.5$, $\eta=0.1$.
 (a1) corresponds to the longitudinal
  pressure (\ref{pl}),  (a2) corresponds to the
  transverse pressure (\ref{pt}),  the dashed curve (a3) corresponds
  to the transverse pressure in the absence of dust ($\eta=0$),
  (b1,b2) correspond to the electric field for $\eta\neq0$ and
  $\eta=0$.
  The vertical dashed lines in both figures show the position of the
   sheath-presheath interface
   ($z_s=10.31$).
  Discontinuities at the curves (a1, a2, b1) are at the equilibrium
  position of the dust layer ($z_l=3.94$).
  }
\end{figure}

Before discussing the results of the numeric evaluation of the
surface tension, it is convenient to introduce the normalized
quantities. In the following, the coordinate, $z$, is measured in
the units of the electron Debye length, $\lambda_D=\sqrt{T_e/4\pi
e^2 n_0}$, the
 pressure is normalized to $n_0 T_e$, and
 the surface tension is normalized to $\lambda_D n_0 T_e$.
 The dimensionless surface
 density of the dust layer is $\eta=q\sigma /(2 e n_0 \lambda_D)$
 and the normalized grain mass is $\mu=M g e \lambda_D/(q T_e)$.
 Under the typical experimental conditions,   $\mu \leq 1$,
 while $\eta \leq 0.1$. All computations discussed below have been
 performed with the sufficiently small value of the ionization
 frequency, $\nu_{ion}/\omega_{pi}=0.01$, where
 $\omega_{pi}$ is the ion plasma
 frequency.

 Examples of the pressure and the electric field profiles are depicted in
 Fig.~\ref{fig2}. The net simulation length for this particular
 run is $z_0=68.5$. In order to make the effect discernible in the
 figure,
 we use here a sufficiently large value of the normalized dust density,
 $\eta=0.1$.  The longitudinal pressure drop (the curve 1 in
 Fig.~\ref{fig2}a) at the charged layer is compensated by its
 weight; outside of the layer, the longitudinal pressure is
 constant.
 Contrary, the transverse pressure (\ref{pt}) is strongly
 inhomogeneous inside the sheath (the curves 2 and 3 in
 Fig.~\ref{fig2}a). Of importance is that
  the  dust reduces the transverse pressure
 between the layer and the electrode. As it is readily seen from
Fig.~\ref{fig2}b,   the effect is mostly due to
 the reduction of the electric field.

  The
 vertical dashed lines in Fig.~\ref{fig2} show the position of the sheath-presheath
 interface, $z_s$, for $\eta\neq0$, which is conditionally defined as the point  where the ion flow
  velocity is equal to the ion sound velocity. The dependence of the coordinate
   $z_s$ on the dust parameters is rather weak. For this
   particular example, the dust layer levitates in the supersonic
   ion flow, $z_l<z_s$; with reducing grain weight, $\mu$, it
   moves up to the subsonic area, $z_l>z_s$.

\begin{figure}
\includegraphics{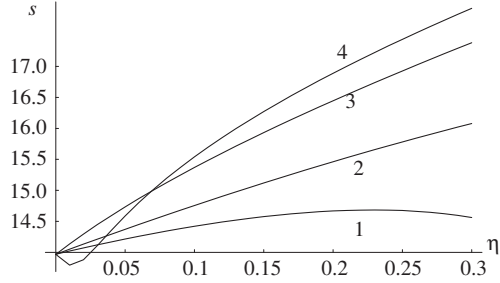}
 \caption{\label{fig3} Dependence of the normalized surface tension on the dust density, $\eta$.
 1 --- $\mu=0.8$, 2 --- $\mu=0.5$, 3 --- $\mu=0.1$, 4 --- $\mu=0.02$. }
\end{figure}

\begin{figure}
\includegraphics{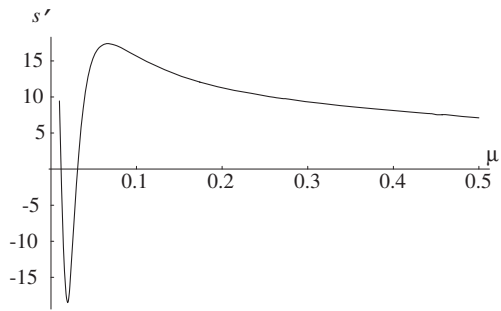}
 \caption{\label{fig4} The derivative of the surface tension with respect
 to the dust density versus the grain mass. }
\end{figure}

Fig.~\ref{fig3} shows the dependence of the surface tension on the
dust layer density, $\eta$, for various values of the grain mass,
$\mu$. The curves 1-3 in Fig.~\ref{fig3} correspond to the dust
layer levitating in the supersonic flow, $z_l<z_s$,  the curve 4
corresponds to lighter dust grains situated in the presheath area,
$z_l>z_s$. The depicted curves show that if the dust layer is
located in the sheath, the surface tension increases with $\eta$,
while with the dust layer shifted to the presheath, the surface
tension may become a decreasing function of $\eta$.  This is also
shown in Fig.~\ref{fig4}, where the dependence of the derivative
$s^\prime(\eta)|_{\eta=0}$ on the grain mass, $\mu$, is plotted.

As we have already mentioned, the dependence of the surface
tension on the dust density determines the force acting upon the
edge of the semi-infinite layer. If $s(\eta)>s(0)$ then the force
$f_x=s(0)-s(\eta)$ is negative, as shown in Fig.~\ref{fig1}, that
is, the surface tension at the plasma-wall interface tends to
expel the dust. Otherwise, the dust is drawn in the sheath. As it
is seen from  Figs.~\ref{fig3},\ref{fig4}, in dependence on the
grain mass and the dust density both alternatives are possible.

Here, we treat dust layer as a rigid incompressible medium.
However, the short-range interaction between the dust grains
result in the surface pressure, $p_s(\sigma)$, which also depends
on the density \cite{hebner}. The surface tension (\ref{ps})
reduces the surface pressure, that is, the net equation of state
of the dust layer is given by $p=p_s(\sigma)-s(\sigma)$. Since
$p_s(\sigma)$ exponentially tends to zero with decreasing
$\sigma$, the compressibility, $dp/d\sigma$,   of a rarefied dust
layer is layer is negative if $s^\prime(0)>0$. Such a medium is
unstable. The detailed investigation of this instability and
arising structures will be discussed elsewhere.

To summarize, charged dust may be treated as  a surface-active
substance that is capable of altering the surface tension  at the
plasma-electrode interface. Unlike usual surfactants, the dust
increases the surface tension in a wide range of parameters.
Although we did not discuss here the emergence of the
two-dimensional void \cite{tue}, it should be pointed out that
 a similar effect is
well-known in liquids. To observe a void, it is enough to  drop
some liquid soap at the surface of water with floating fine
powder.

 This study was supported in part by the Netherlands
Organization for Scientific Research (grant no. NWO 047.008.013)
and the Russian Foundation for Basic Research (project no.
02-02-16439).

\end{document}